\documentclass{PoS}

\usepackage{amsmath}

\title{Linearly polarised Transverse Momentum Dependent Parton Distribution Function at NNLO in QCD.}

\ShortTitle{Linearly polarised gluons at NNLO in QCD}

\author{\speaker{Sergio Leal G\'omez}$^{a,b}$\\
        \llap{$^a$}Departamento de F\'isica Te\'orica and IPARCOS,
        Universidad Complutense de Madrid (UCM),
        28040 Madrid, Spain.\\
        \llap{$^b$}University of Vienna, Faculty of Physics, Boltzmanngasse 5, A-1090 Wien, Austria.\\
        E-mail: \email{sergiole@ucm.es}}

\author{Daniel Gutierrez-Reyes$^{a}$\\
        \llap{$^a$}Departamento de F\'isica Te\'orica and IPARCOS,
        Universidad Complutense de Madrid (UCM),
        28040 Madrid, Spain.\\
        E-mail: \email{dangut01@ucm.es}}
    
\author{Ignazio Scimemi$^{a}$\\
	\llap{$^a$}Departamento de F\'isica Te\'orica and IPARCOS,
	Universidad Complutense de Madrid (UCM),
	28040 Madrid, Spain.\\
	E-mail: \email{ignazios@ucm.es}}

\author{Alexey Vladimirov$^{c}$\\
	\llap{$^c$}Institut f\"ur Theoretische Physik, Universit\"at Regensburg,
	D-93040 Regensburg, Germany.\\
	E-mail: \email{alexey.vladimirov@ur.de}}

\abstract{We present for first time the computation at large $q_T$ (or small $b_T$) matching coefficients of Transverse Momentum Dependent Parton Distribution Function (TMDPDF) for linearly polarised to the integrated gluon distribution at next-to-next-to leading order (NNLO). The computation is performed using the modified $\delta$-regulator for rapidity divergences and dimensional regularization. This TMDPDF matriz element presnts phenomenological interest in two kind of processes. The factorization of the Higgs production transverse momentum ($q_T$) distribution through gluon-gluon fusion and the quarkonium production.}

\FullConference{XXVII International Workshop on Deep-Inelastic Scattering and Related Subjects - DIS2019\\
		8-12 April, 2019\\
		Torino, Italy}

\begin{document}

\section{Introduction}

The transverse momentum dependent (TMD) factorisation theorems for semi-inclusive deep inelastic scattering (SIDIS) and Drell-Yan type processes formulated in \cite{GarciaEchevarria:2011rb,Collins:2011zzd,Chiu:2012ir, Echevarria:2014rua} allow a consistent treatment of the rapidity divergences in the definition of spin (in)dependent TMD distributions. Within factorisation theorems TMD operators are self-contained defined objects, and can be considered individually by standard methods of quantum field theory without referring to a scattering process. TMD operators are intrinsically non-perturbative objects due to the infrared divergences, nevertheless in the limit of large-$q_t$ (or small $b_T$) a perturbative computation of matching coefficients of TMD distributions on the corresponding integrated functions can be performed. The unpolarised TMD distribution is the most studied case and it has been treated using different regularization schemes at the next-to-leading order (NLO) \cite{GarciaEchevarria:2011rb,Collins:2011zzd,Echevarria:2014rua} and the next-to-next-to-leading order (NNLO) \cite{Echevarria:2015usa,Echevarria:2016scs,Gehrmann:2014yya,Gehrmann:2012ze}. For polarised distribution such a program has been performed for helicity, transversity, pretzelosity and linearly polarised distribution at NLO \cite{Bacchetta:2013pqa,Echevarria:2015uaa,Gutierrez-Reyes:2017glx} and only for transversity and pretzelosity at NNLO \cite{Gutierrez-Reyes:2018iod}. In this work we present the TMDPDF for linearly polarised gluons at NNLO improving the status of the art on polarised TMD distributions. The phenomenological interest of gluons TMD distribution is supported on the fact that at high energies most of the scattering processes are triggered by gluons. 

Gluon-gluon fusion is the main channel for Higgs production. The factorisation of this process in the infinite top-mas ($m_t$) limit and with $q_T\ll m_H$, where $q_T$ is the transverse momentum spectrum of the Higgs boson produced via gluon-gluon fusion and $m_H$ is its mass, has been demonstrated to follow the same pattern as in the Drell-Yan/vector boson case, and in this sense it has been reviewed in \cite{Echevarria:2015uaa}. We find intriguing the fact that the sign of linearly polarised gluon contribution can flip depending on the (pseudo) scalar nature of the Higgs boson \cite{Boer:2011kf, Boer:2013fca}. The possibility to test the parity of the Higgs relies heavily on the precision achievable experimentally, and a theoretical prediction which includes resummation at next-to-leading logarithmic (NLL) order has been done in \cite{Boer:2014tka}.

Another process where appears linearly polarised gluons is in the di-$J/\psi$ production, which leads a modulation $\cos\left(2\phi\right)\left(\cos\left(4\phi\right)\right)$ in the azimuthal angle due to single(double) gluon helicity flips  \cite{Lansberg:2017dzg}.

\section{Matching Coefficients}

In this section we define the basic operators for the case of interesting following \cite{Gutierrez-Reyes:2017glx}, and show the matching coefficients, which are the main result of this work. The gluon  TMD operator matrix element reads\footnote{We omit the transverse links necessary in singular gauges.}
\begin{equation}
\begin{aligned}
	\Phi_{g\leftarrow h, \mu\nu}(x,\vec b)=&\langle P,S|\frac{1}{xp^+}\int \frac{d\lambda}{2\pi}e^{-ixp^+\lambda}
	\bar T \left\{F_{+\mu}(\lambda n+\vec b)
	\tilde W_{n}(\lambda n+\vec b)\right\} 
    \\ &\times
	T\left\{ \tilde W_{n}^{\dagger}(0)
	(\lambda,\vec b)F_{+\nu}(0)\right\}|P,S\rangle,
\end{aligned}
\label{def:TMD_OP_G}
\end{equation}
where $n$ is the lightlike vector and we use the standard notation for the lightcone components of vector $v^\mu=n^\mu v^-+\bar n^\mu v^++g_T^{\mu\nu}v_\nu$ (with $n^2=\bar n^2=0$, $n\cdot\bar n=1$, and $g_T^{\mu\nu}=g_{\mu\nu}-n^\mu \bar n^\nu-\bar n^\mu n^\nu$). 
The Wilson lines $\tilde W$ are taken in the adjoint representation of the gauge group for the gluon case. 
The hadron matrix elements of the TMD operators in eq.~(\ref{def:TMD_OP_G}) are decomposed in covariant Lorentz structures, the TMDPDF.
The decomposition of gluon operator in momentum space over all possible Lorentz variants can be found in \cite{Mulders:2000sh}.
Here we use the corresponding decomposition in impact parameter space which is more convenient when treating the formulation of the factorisation theorem.
The corresponding between decomposition in momentum and impact parameter space can be found in e.g. \cite{Boer:2011xd,Echevarria:2015uaa}.
In $b$-space we have

\begin{equation}
\begin{aligned}
\Phi_{g\leftarrow h}^{\mu\nu}(x,\vec b)=&
\frac{1}{2}\Big(-g_T^{\mu\nu}f_{1,g\leftarrow h} (x,\vec b)-
i\epsilon_T^{\mu\nu}S_Lg_{1L, g\leftarrow h} (x,\vec b)\\
&+2h_{1,g\leftarrow h}^{\perp} (x,\vec b)(\frac{g_T^{\mu\nu}}{2}+\frac{ b^\mu  b^\nu}{\vec b^2})+...\Big),
\end{aligned}
\label{TMD_G_dec}
\end{equation}
where the vector $b^\mu$ is a 4-dimensional vector of the impact parameter ($b^+=b^-=0$ and $-b^2\equiv\vec{b}^2>0$). On the r.h.s of eq. ~(\ref{TMD_G_dec}) $f_1^g$ is the unpolarised gluon TMDPDF, $g_{1L}^g$ the helicity and finally we have $h_1^{\perp g}$ as the TMD for linearly polarised gluons, which is the object of the present work.
In eq. ~(\ref{TMD_G_dec}) we write only the TMD distribution that match the twist-2 integrated parton distribution function (PDF) and the twist-3 and higher parton distribution function are understood in the dots.

The small-$b$ operator product expansion (OPE) provides a relation between a TMD operators and collinear integrated operators which at lower twist reads
\begin{equation}
\Phi_{g\leftarrow h, \mu\nu}(x,\vec b)=\sum_f C_{g\leftarrow f, \mu\nu}^a(x,\vec b)\otimes \phi_{f\leftarrow h}^{a, {\rm{tw2}}}(x)+\dots
\label{OPE_G}
\end{equation}
where symbol $\otimes$ denotes the Mellin convolution in the variable $x$, the function $C(x,\vec{b})$ depend on $\vec{b}$ only logarithmically, $\phi^a\left(x\right)$ are collinear functions, the dots represent the power suppressed contributions and scale dependences are not shown.

at the lowest order of PDF, the function $\phi\left(x\right)$ are the formal limit of the TMD distribution $\Phi(x,\vec{0})$. The hadronic matrix elements of $\phi$ are the gluon/quark PDFs.

It is also convenient to extract to desired TMD using projectors and defining
\begin{eqnarray}
\Phi_{g\leftarrow h}^{[\Gamma]}=\Gamma^{\mu\nu}\Phi_{g\leftarrow h,\mu\nu}.
\end{eqnarray}
The projector for unpolarised gluons is
\begin{equation}
\Gamma^{\mu\nu}_{un}=\frac{g^{\mu\nu}_T}{2(1-\epsilon)}
\end{equation}
and for linearly polarised gluons we have
\begin{equation}
\Gamma^{\mu\nu}_{\ell in}=\left(g^{\mu\nu}_T-2 (1-\epsilon) \frac{b^\mu b^\nu}{b^2} \right)\frac{1}{2(1-2\epsilon)}.
\label{eq:l-proj}
\end{equation}
The small-$b$ matching of this distribution has been performed  in \cite{Echevarria:2015uaa,Gutierrez-Reyes:2017glx} up to NLO.

The structure of rapidity divergences for the gluon TMD operators differs from the quark case only because of the colour factors. As for all TMDs, both ultraviolet (UV) and rapidity divergences are present in their perturbative calculation which are renormalized by appropriate renormalisation constant \cite{Echevarria:2016scs,Vladimirov:2017ksc}.
Hence, the renormalisation (or physical) TMDPDF depend on two scales (the UV renormalisation scale is denoted by 
$\mu$ and the rapidity renormalisation scale is denoted by $\zeta$). The renormalised expression for gluon TMDPDF and renormalisation constan can be found in \cite{Echevarria:2016scs}.

In  perturbation theory, the expression for the coefficient function can be presented as
\begin{equation}
\delta^L C_{g\leftarrow f'}(x,\mathbf{L}_\mu,\mathbf{l}_\zeta)=\sum_{n=0}^\infty a_s^n \sum_{k=0}^{n+1}\sum_{l=0}^n \mathbf{L}_\mu^k\, \mathbf{l}_\zeta^l \, \delta^L C^{(n;k,l)}_{g\leftarrow f'}(x),
\label{eq:pertTR}
\end{equation}
where $a_s=g^2/(4\pi)^2$. The coefficients $\delta^L C^{(n;k,l)}$ with $k+l>0$ are  fixed order-by-order  with the help of the renormalisation group equations and they can be found up to two loops in  e.g. \cite{Echevarria:2015usa} as they are common to the unpolarised case.

The only non-trivial part up to two loops is so provided by  $\delta^L C^{(2;0,0)}$, where

\begin{equation}
\begin{aligned}
\textcolor{black}{\delta^L C_{g\leftarrow g}^{(2;0,0)}(x)}& \textcolor{black}{=C_A^2\Big\{\frac{1}{x}(\frac{220}{9}+20(1-x)\zeta_2+16\ln x)-32 \frac{1-x}{x}{\rm Li}_3(1-x)} \\
&\textcolor{black}{-[16\frac{1-x}{x}+\frac{x}{2}(x+3)]{\rm Li}_2(x)} \\
&\textcolor{black}{+\frac{x+3}{4}[x{\rm Li}_3(x^2)+(1-x\ln x){\rm Li}_2(x^2)]} \\
&\textcolor{black}{+\frac{x+3}{2}\ln x[(1-x)\ln(1-x)+\ln(1+x)+\frac{x}{2}]} \\
&\textcolor{black}{+\frac{76}{3}\ln x+x(\frac{77}{6}+\frac{31}{36}x)-8\ln^2 x} \\ 
&\textcolor{black}{+\frac{x}{4}(x+3)(\zeta_2-\zeta_3)-\frac{1325}{36}\Big\}} \\
&\textcolor{black}{+C_F T_R n_f\Big\{8 \ln^2x-16\frac{(1-x)^3}{x} \Big\}} \\
&\textcolor{black}{+C_A T_R n_f\Big\{\frac{136}{9x} +\frac{16}{3}\ln x-\frac{8 x}{9}(x+3)-\frac{128}{9}\Big\},}
\end{aligned}
\label{C2gg}
\end{equation}

\begin{equation}
\begin{aligned}
\textcolor{black}{
	\delta^L C^{(2;0,0)}_{g\leftarrow q}(x)}&\textcolor{black}{=C_F C_A \Big\{-16 \frac{1-x}{x}[{\rm Li}_2(x)+2{\rm Li}_3(1-x)]-\frac{40}{3}\frac{1-x}{x}\ln(1-x)}\\
&\textcolor{black}{-8\frac{1-x}{x}\ln^2(1-x)+(40+\frac{16}{x})\ln x-8\ln^2 x}\\
&\textcolor{black}{+\frac{1-x}{x}(\frac{88}{9}+20\zeta_2-8x)\Big\}}\\
&\textcolor{black}{+C_F^2\Big\{8(1-x)+4(\ln x-5)\ln x+8\frac{1-x}{x}[1+\ln(1-x)]\ln (1-x)\Big\}}\\
&\textcolor{black}{+\frac{32}{3} C_F T_R n_f\frac{1-x}{x}(\ln(1-x)+\frac{2}{3}).}
\end{aligned}
\label{C2gq}
\end{equation}
Note that we have no singularity for $x\rightarrow 1$.

\section{Conclusions}

We have provided a description of TMDPDF for linearly polarised at the same order of precision than TMDPDF for unpolarised gluons. We find out that TMDPDF for linearly polarised gluons is not divergent when x goes to 1, unlike for unpolarised gluons. 

We have reviewed that the contribution of linearly polarised gluons to the Higgs cross section through gluon-gluon fusion is less than \%. This can be understood due to the fact that Higgs production takes place at electroweak energies scale, together to the non divergent behaviour of linearly polarised gluons when x $\rightarrow 1$ makes the contribution of unpolarized gluons to exceed linearly polarised gluons.

For quarkonium production, di-$J/\psi$ production is the most promising process to measure linearly polarised gluons effects, which effect can reach the 50\% of the corss section. Recently quarkonium production has been factorised \cite{Echevarria:2019ynx} which leads   interesting opportunities to work in quarkonium production phenomenology.

\section*{Acknowledgements}
D.G.R., S.L.G. and  I.S. are supported by the Spanish MECD grant FPA2016-75654-C2-2-P. 
This project has received funding from the European Union Horizon 2020 research and innovation program under grant agreement No 824093 (STRONG-2020). S.L.G is supported by the Austrian Science Fund FWF under the Doctoral Program W1252-N27 Particles and Interactions.

\bibliographystyle{JHEP}  
\bibliography{TMD_ref}

\end{document}